\documentclass[12pt,preprint]{aastex}

\usepackage {psfig}
\slugcomment{}

\shorttitle{The Angular Correlation Function}

\author{Andrew J. Connolly\altaffilmark{1},
Ryan Scranton\altaffilmark{2,3}, David Johnston\altaffilmark{2,3},
Scott Dodelson\altaffilmark{2,3},
Daniel J. Eisenstein\altaffilmark{2,4,27}, Joshua A. Frieman\altaffilmark{2,3},
James E. Gunn\altaffilmark{5},
Lam Hui\altaffilmark{6},
Bhuvnesh Jain\altaffilmark{7},
Stephen Kent\altaffilmark{3},
Jon Loveday\altaffilmark{8}, 
Robert C. Nichol\altaffilmark{9}, Liam O'Connell\altaffilmark{8},
Marc Postman\altaffilmark{10},
Roman Scoccimarro\altaffilmark{11,12},
Ravi K. Sheth\altaffilmark{3}, Albert Stebbins\altaffilmark{3},
Michael A. Strauss\altaffilmark{5}, Alexander S. Szalay\altaffilmark{13},
Istv\'an Szapudi\altaffilmark{14}, Max Tegmark\altaffilmark{7},
Michael S. Vogeley\altaffilmark{15},
Idit Zehavi\altaffilmark{3},
James Annis\altaffilmark{3},
Neta Bahcall\altaffilmark{5},
J. Brinkmann\altaffilmark{16},
Istv\'an Csabai\altaffilmark{17},
Mamoru Doi\altaffilmark{18},
Masataka Fukugita\altaffilmark{19},
G.S. Hennessy\altaffilmark{20},
Robert Hindsley\altaffilmark{21},
Takashi Ichikawa\altaffilmark{22},
\v{Z}eljko Ivezi\'c\altaffilmark{5},
Rita S.J. Kim\altaffilmark{5},
Gillian R. Knapp\altaffilmark{5}, 
Peter Kunszt\altaffilmark{13}, 
D.Q.\ Lamb\altaffilmark{2}, 
Brian C. Lee\altaffilmark{3},
Robert H. Lupton\altaffilmark{5}, 
Timothy A. McKay\altaffilmark{23},
Jeff Munn\altaffilmark{24},
John Peoples\altaffilmark{3},
Jeff Pier\altaffilmark{24}, Constance Rockosi\altaffilmark{2},
David Schlegel\altaffilmark{5}, Christopher Stoughton\altaffilmark{3},
Douglas L. Tucker\altaffilmark{3},
Brian Yanny\altaffilmark{3}, Donald G. York\altaffilmark{2,25}, 
for the SDSS Collaboration
}
 
\altaffiltext{1}{University of Pittsburgh, Department of Physics and
Astronomy, Pittsburgh, PA 15260, USA}
\altaffiltext{2}{Astronomy and Astrophysics Department, University of
Chicago, Chicago, IL 60637, USA}
\altaffiltext{3}{Fermi National Accelerator Laboratory, P.O. Box 500, 
Batavia, IL 60510, USA}
\altaffiltext{4} {Steward Observatory, University of Arizona, 933 N.\ Cherry Ave., Tucson, AZ 85721}
\altaffiltext{5}{Princeton University Observatory, Princeton, NJ 08544,
USA}         
\altaffiltext{6}{Department of Physics, Columbia University, New York, NY
10027, USA}
\altaffiltext{7}{Department of Physics, University of Pennsylvania,
Philadelphia, PA 19101, USA}
\altaffiltext{8}{Sussex Astronomy Centre, University of Sussex, Falmer,
Brighton BN1 9QJ, UK}
\altaffiltext{9}{Department of Physics, 5000 Forbes Avenue, Carnegie
Mellon University, Pittsburgh, PA 15213, USA}
\altaffiltext{10}{Space Telescope Science Institute, Baltimore, MD 21218, USA}
\altaffiltext{11}{Department of Physics, New York University, 4 Washington
Place, New York, NY 10003}
\altaffiltext{12}{Institute for Advanced Study, School of Natural
Sciences, Olden Lane, Princeton, NJ 08540, USA}
\altaffiltext{13}{Department of Physics and Astronomy, The Johns Hopkins
University,  Baltimore, MD 21218, USA}
\altaffiltext{14}{Institute for Astronomy, University of Hawaii, 2680
Woodlawn Drive, Honolulu, HI 96822, USA}
\altaffiltext{15}{Department of Physics, Drexel University, Philadelphia,
PA 19104, USA}
\altaffiltext{16}{Apache Point Observatory, P.O. Box 59, Sunspot, 
NM 88349-0059}
\altaffiltext{17}{Department of Physics of Complex Systems,E\"{o}tv\"{o}s 
University, Budapest, Hungary, H-1088}  
\altaffiltext{18}{University of Tokyo, Institute of Astronomy and 
Research Center of the Early Universe, School of Science}
\altaffiltext{19}{University of Tokyo, Institute for Cosmic Ray Research, 
Kashiwa, 2778582}
\altaffiltext{20}{US Naval Observatory, 3450 Massachusetts Ave., NW, Washington, DC 20392-5420}
\altaffiltext{25}{Remote Sensing Division, Naval Research 
Laboratory, 45555 Overlook Ave. SW, Washington DC 20375}
\altaffiltext{22}{Astronomical Institute, Tohoku University, Aoba, Sendai 980-8578, Japan}
\altaffiltext{23}{University of Michigan, Department of Physics, 
500 East University, Ann Arbor, MI, 48109}
\altaffiltext{24}{US Naval Observatory, Flagstaff Station, P.O.\ Box 1149, Flagstaff, AZ 86002-1149}
\altaffiltext{25}{Enrico Fermi Institute, 5640
South Ellis Avenue, Chicago, 60637}
\altaffiltext{26}{Hubble Fellow}

\newcommand{\etal}{{\it et al.}}

\begin{document}

\title{The Angular Correlation Function of Galaxies from Early SDSS Data}

\begin{abstract}
The Sloan Digital Sky Survey is one of the first multicolor
photometric and spectroscopic surveys designed to measure the
statistical properties of galaxies within the local Universe. In this
Letter we present some of the initial results on the angular 2-point
correlation function measured from the early SDSS galaxy data.  The
form of the correlation function, over the magnitude interval
$18<r^*<22$, is shown to be consistent with results from existing
wide-field, photographic-based surveys and narrower CCD galaxy
surveys. On scales between 1 arcminute and 1 degree the correlation
function is well described by a power-law with an exponent of $\approx
-0.7$. The amplitude of the correlation function, within this angular
interval, decreases with fainter magnitudes in good agreement with
analyses from existing galaxy surveys. There is a characteristic break
in the correlation function on scales of approximately 1--2
degrees. On small scales, $\theta <$ 1', the SDSS correlation function
does not appear to be consistent with the power-law form fitted to the
$1'<\theta<0.5^\circ$ data.

With a data set that is less than 2\% of the full SDSS survey area, we
have obtained high precision measurements of the power-law angular
correlation function on angular scales $1' < \theta < 1^\circ$, which
are robust to systematic uncertainties. Because of the limited area
and the highly correlated nature of the error covariance matrix, these
initial results do not yet provide a definitive characterization of
departures from the power-law form at smaller and larger angles. In
the near future, however, the area of the SDSS imaging survey will be
sufficient to allow detailed analysis of the small and large scale
regimes, measurements of higher-order correlations, and studies of
angular clustering as a function of redshift and galaxy type.

\end{abstract}


\section{Introduction}

From the earliest wide field photographic-plate surveys (Shane \&
Wirtanen 1967) the statistics of point processes (Totsuji \& Kihara
1969) have been used to characterize the angular clustering of
galaxies and its evolution as a function of magnitude and
redshift. One of the simplest of these statistics is the 2-point
correlation function, which measures the excess of pairs of galaxies
as a function of separation when compared to a random
distribution. For the case of a Gaussian random field, the 2-point
correlation function and its Fourier transform pair, the power
spectrum, provide a complete representation of the statistical
fluctuations in the distribution of galaxies. Even for the case of
non-Gaussianity, the 2-point function provides a simple and important
statistical test of galaxy formation models.

At bright magnitudes, the angular 2-point function has been studied
from large galaxy surveys such as the Lick survey (Groth \& Peebles
1977), the Automated Plate Measuring galaxy survey (APM; Maddox {\it
et al.}  1990), and the Edinburgh Durham Southern Galaxy Catalogue
(EDSGC; Collins {\it et al.} 1992), which cover several thousand
square degrees of the sky. From these surveys the correlation function
has been characterized on scales of a few tens of arcseconds to
several degrees (corresponding to physical scales up to about 15
$h^{-1}$ Mpc.) The correlation function, within these galaxy surveys,
is consistently found to be a power law on small scales, with a break
at approximately 2 degrees (for a survey with an apparent magnitude
limit of $r^*=20$), beyond which it falls off more steeply.  On scales
smaller than the break, the exponent of the power law, defined by
$w(\theta) = A\theta^{1-\gamma}$, has a value $\gamma\approx1.7$
(Peebles 1980).  Deeper surveys, either from photographic plates or
CCD cameras, show that the power-law form of the small-scale
correlation function remains with the amplitude decreasing with 
fainter magnitudes (Stevenson et al. 1985, Couch {\it et al.} 1993,
Roche {\it et al.} 1993, Hudon \& Lilly 1996, Postman {\it et al.}
1998).

While these surveys have provided a powerful description of the
angular clustering of galaxies, they were either based on visual or
machine scans of photographic plates or were limited to a relatively
small areal coverage.  In this paper, we report the first results from
a series of studies of the angular clustering of galaxies from the
commissioning data of the Sloan Digital Sky Survey (SDSS; Stoughton
{\it et al.} 20001, York {\it et al.}  2001).  This represents the
first systematic, wide-field, CCD survey, specifically designed to
measure the statistical properties of galaxies within the local
Universe.  The present results use data from 160 square degrees. In
the following sections, we describe the imaging data used to construct
the angular correlation function and the statistical analyses applied
to these data.  We compare the resultant angular correlation functions
with those published in the literature and show that even with less
than 2\% of the final survey data, the clustering statistics are of
comparable precision to those previously published.

In a series of companion papers to this short communication, led by
the crucial precursor paper that provides the detailed tests for
systematic effects within the photometric data \citep{scranton01}, we
discuss a measurement of the galaxy angular power spectrum that
complements our $w(\theta)$ measurements by better characterizing
fluctuations on large scales \citep{tegmark01}. An inversion of the
angular correlation function and angular power spectrum to infer the
3-dimensional galaxy power spectrum is discussed by Dodelson {\it et
al.} 2001.  \citet{szalay01} use the angular galaxy distribution to
constrain power spectrum parameters by a maximum-likelihood technique.
A first analysis of clustering in the SDSS redshift survey appears in
\citet{zehavi01}.  Future papers will examine higher order
correlations in the angular and redshift data \citep{szapudi01} and
the impact of gravitational lensing on the galaxy angular correlation
function \citep{jain01}.

\section{The Sloan Digital Sky Survey}

The Sloan Digital Sky Survey is a wide-field photometric and
spectroscopic survey being undertaken by the Astrophysical Research
Consortium at the Apache Point Observatory in New Mexico (York {\it et
al.} 2000).  The completed survey will cover approximately 10,000
square degrees.  CCD imaging with the SDSS camera (Gunn \etal\ 1998)
will image $10^8$ galaxies in five colors ($u^*$, $g^*$, $r^*$, $i^*$,
and $z^*$; see Fukugita \etal\ 1996) to a detection limit,of
approximately $r^*=23$ at $5:1$ signal-to-noise ratio. From these
data, 1 million galaxies will be selected for spectroscopic followup
(Strauss {\it et al.} 2001, Eisenstein {\it et al.} 2001). In this
Letter, we focus on the analysis of a small subset of the final survey
volume that were taken during the commissioning of the survey
telescope in 1999.

The results presented here are based on two nights of imaging data
(designated Runs 752 and 756) taken on the 20th and 21st of March
1999. These interleaved scans are centered on the Celestial Equator
covering a $2.5^\circ$ stripe with declination $|\delta|<1.25^\circ$
and Right Ascension ranging from $7^h\,49^m$ to $16^h\,46^m$.  Within
this stripe, regions of data where the imaging quality had seeing
values better than $1.75"$ (as measured in the $r^*$ band) were
extracted and a Bayesian star-galaxy separation algorithm applied
(Scranton {\it et al.} 2001, Lupton {\it et al.} 2001b).  The
photometric pipeline (Lupton {\it et al.} 2001a) was used to calculate
magnitudes for each object based on the best-fit, PSF-convolved de
Vaucouleurs or exponential model, including an arbitrary scale size
and axis ratio.  All magnitudes for sources within the high quality
subset were corrected for Galactic extinction using the reddening maps
of Schlegel, Finkbeiner \& Davis (1998).  In total, the final catalog
contains $1.46\times 10^6$ galaxies in the apparent magnitude range of
$18<r^*<22$ and covers an area of 160 square degrees. These data
comprise a subset of the data made publicly available as part of the
Early Data Release of the SDSS (Stoughton \etal\ 2001).

The uniqueness of the SDSS data lies in its uniformity. The angular
clustering signal on large scales is small: at one degree, the
amplitude of $w(\theta)$ is approximately 0.006 for a magnitude range
of $19<r^*<20$. Systematic fluctuations of the density of galaxies
due, for example, to errors in the photometric calibration as a
function of position on the sky could dominate the clustering
signal. By controlling these sources of systematic error we have the
opportunity to measure the large scale angular clustering signal. For
the current estimates of photometric uncertainties present in the SDSS
photometry ($\approx 3$\%), fluctuations across the survey volume
might add $\approx0.001$ to the amplitude of $w(\theta)$ (Nichol \&
Collins 1993). These values compare favorably to the errors estimated
from photographic surveys. In Scranton {\it et al.} (2001) we have
performed an extensive search for systematic errors from not only
photometric errors, but also from stellar contamination, seeing,
extinction, sky brightness, bright foreground objects, and optical
distortions in the camera itself. In each case the systematic
uncertainties were found to be small, and in some cases could be
corrected for.

\section{The Angular Correlation Function on Small and Large Scales}

\subsection{Angular 2pt Function Estimators}

The angular correlation function, $w(\theta)$, is calculated from the
estimator of Landy \& Szalay (1993),
\begin{equation}
w(\theta) = \frac{DD - 2DR + RR}{RR},
\end{equation}
where $DD$, $DR$ and $RR$ are pair counts in bins of $\theta\pm \delta
\theta$ of the data-data, data-random and random-random points
respectively. In the limit of weak clustering this statistic is the
2-point realization of a more general representation of edge-corrected
n-point correlation functions (Szapudi \& Szalay 1998) and has been
shown to be close to a minimum variance estimator and to be robust to
the number of random points (Kerscher {\it et al.}  2000).

The correlation function is calculated between 0.001 and 10 degrees
with a logarithmic binning of 6 bins per decade in angle. No integral
constraint correction is applied to these results as the expected
magnitude of this effect is $< 0.0001$ on all scales and magnitude
intervals that are analyzed in this paper (Scranton \etal\ 2001). In
the subsequent analysis we impose a lower limit of 7 arcsec to reduce
artificial correlations due to the decomposition of large galaxies
into multiple sources. At a redshift of $z=0.18$ (the median redshift
of our brightest magnitude shell $18<r^*<19$) this corresponds to
approximately $18 h^{-1}$ kpc. Full details of the analysis of the
variance and covariance in the correlation function, derived from mock
catalogs generated using the PTHalos code (Scoccimarro \& Sheth 2001)
and from jack-knife resampling, are given in Scranton {\it et al.}
(2001). These two approaches give comparable results and for
simplicity we present only the mock catalog errors in the figures.

\subsection{The Angular Correlation function}

In Figure 1a we present the angular correlation function as measured
from the SDSS photometric data over the magnitude interval
$18<r^*<22$. The form of the correlation function is consistent with
that found from extant surveys such as the APM (Maddox {\it et al.}
1990) and EDSGC (Collins {\it et al.} 1992), with a power law on small
scales and a break in the correlation function at approximately 2
degrees.

\begin{figure}[htb]
\plottwo{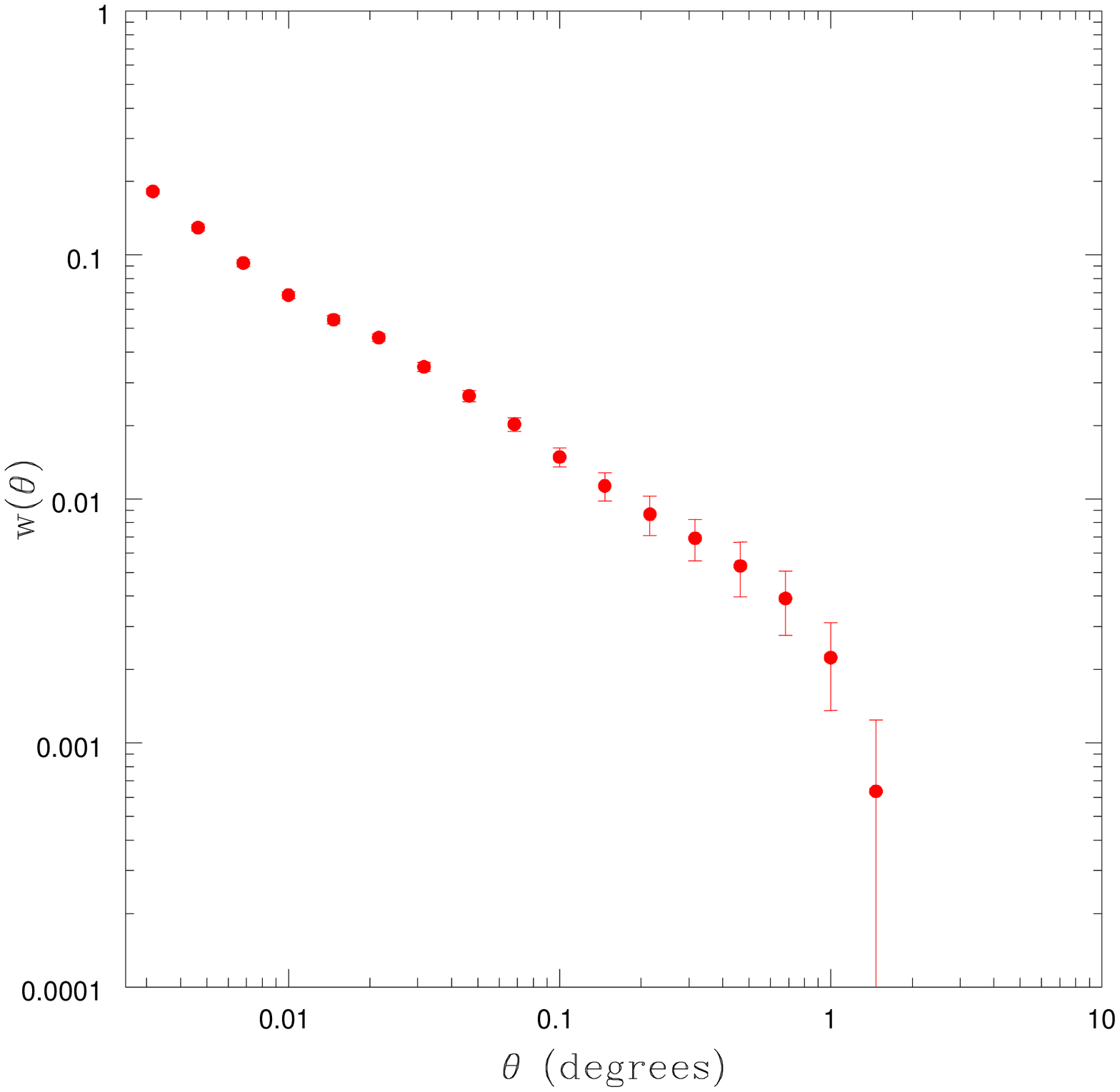}{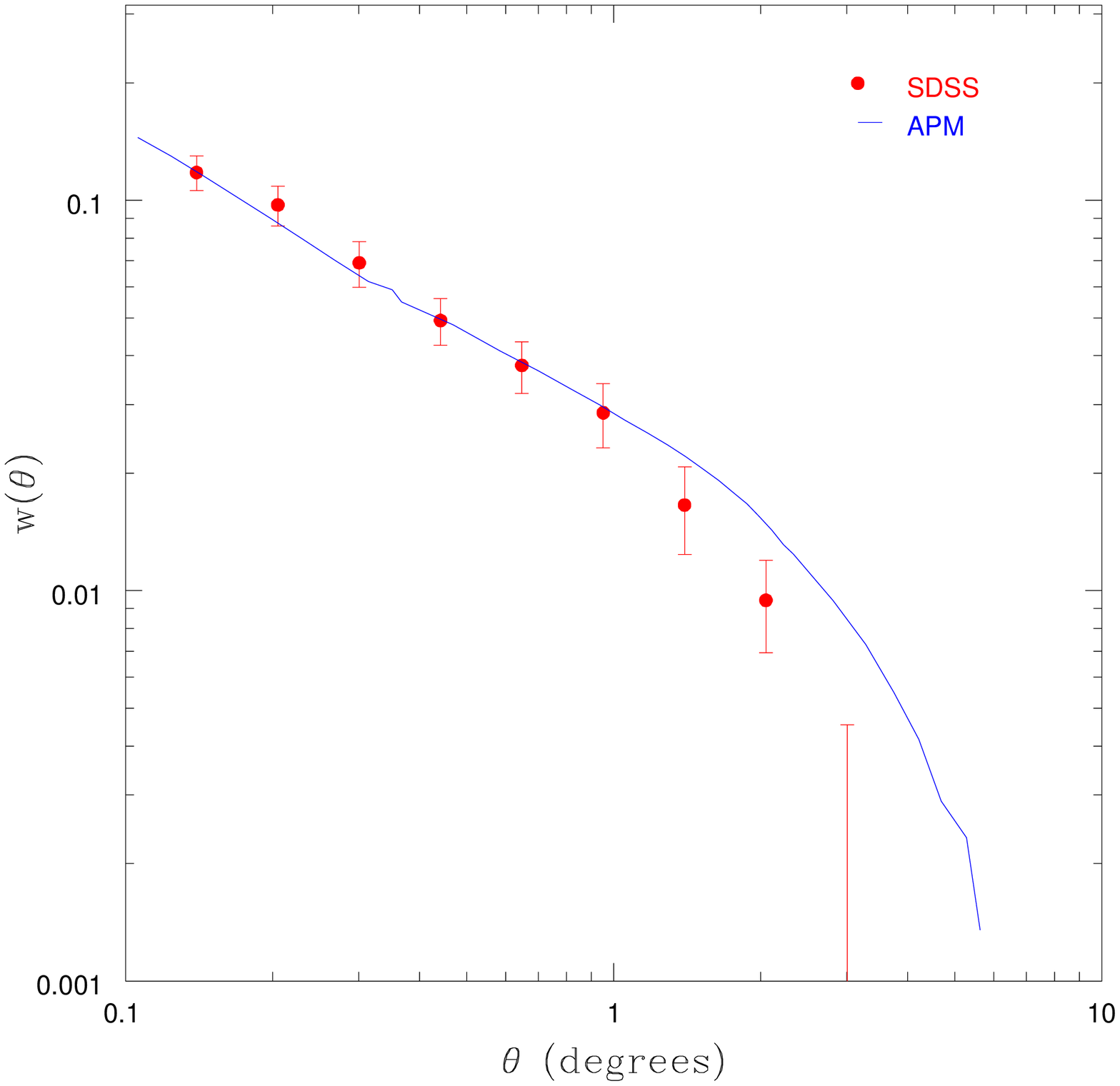}
\label{wm}
\caption{The angular 2-point correlation function from the SDSS early
galaxy data. Figure 1a gives the correlation function measured within
the magnitude interval $18<r^*<22$ (no integral constraint correction
has been applied to these points). The errors on these points are
calculated from jack-knife resampling of the data. In Figure 1b the
SDSS correlation function (filled circles), within the magnitude
interval $18<r^*<19$, is compared with the correlation function from
the APM (Maddox \etal\ 1990) measured over the magnitude interval
$18<B_j<20$ (solid line). The SDSS correlation function has been
scaled to the depth of the APM data using Limber's equation (see text
for details).  }
\end{figure}

In Figure 1b we compare the $w(\theta)$ measurement of the SDSS with
those from the APM galaxy survey.  The solid line represents the
angular correlation function within the magnitude interval $18<B_j<20$
for the APM data and the solid circles represent the correlation
function measured from the SDSS over the magnitude range $18<r^*<19$.
The SDSS correlation function has been scaled to that of the APM using
Limber's equation (Groth \& Peebles 1977), and assuming a cosmological
model with $\Omega=0.3$, $\Omega_{\Lambda}=0.7$, and a double power
law correlation function. The redshift distributions used in the
scaling relation are taken from those presented in Dodelson {\it et
al.}  (2001) for the SDSS data and the empirical redshift relation,
Equation 38, in Maddox {\it et al.}  (1996) for the APM data.  On
scales less than a degree there is good agreement between the
correlation functions.  On scales greater than a degree the amplitude
of the SDSS $18<r^*<19$ subset is lower than the APM correlation
function (though the effect is marginal given the relative errors in
the measurements on these scales).  Given the 160,000 galaxies within
our brightest magnitude slice, the discrepancy between the SDSS and
APM large scale correlation functions can be explained as due purely
to Poisson fluctuations in the number density of galaxies.

\subsection{Angular Correlation Function as a Function of Magnitude}

The scaling of the correlation function with limiting magnitude is
given in Figure 2a for the magnitude intervals $18<r^*<19$, $19<r^*<20$,
$20<r^*<21$ and $21<r^*<22$. The decrease in amplitude of the
correlation function with fainter magnitude slices is consistent with
what we would expect from Limber's equation (Scranton \etal\ 2001).
The solid line shows a power law fit to these data, using the
covariance matrix derived from the mock catalogs, over angular scales
from 1 arcminute to 0.5 degrees.  The application of the full
covariance matrix to these fits is important since neighboring angular
data points are highly correlated.  The exponents, $1-\gamma$, and
amplitudes at 1$^\circ$, A$_w$, of these fits are given in Table 1
together with the $\chi^2$ per degree of freedom of the fits.

\begin{figure}[htb]
\plottwo{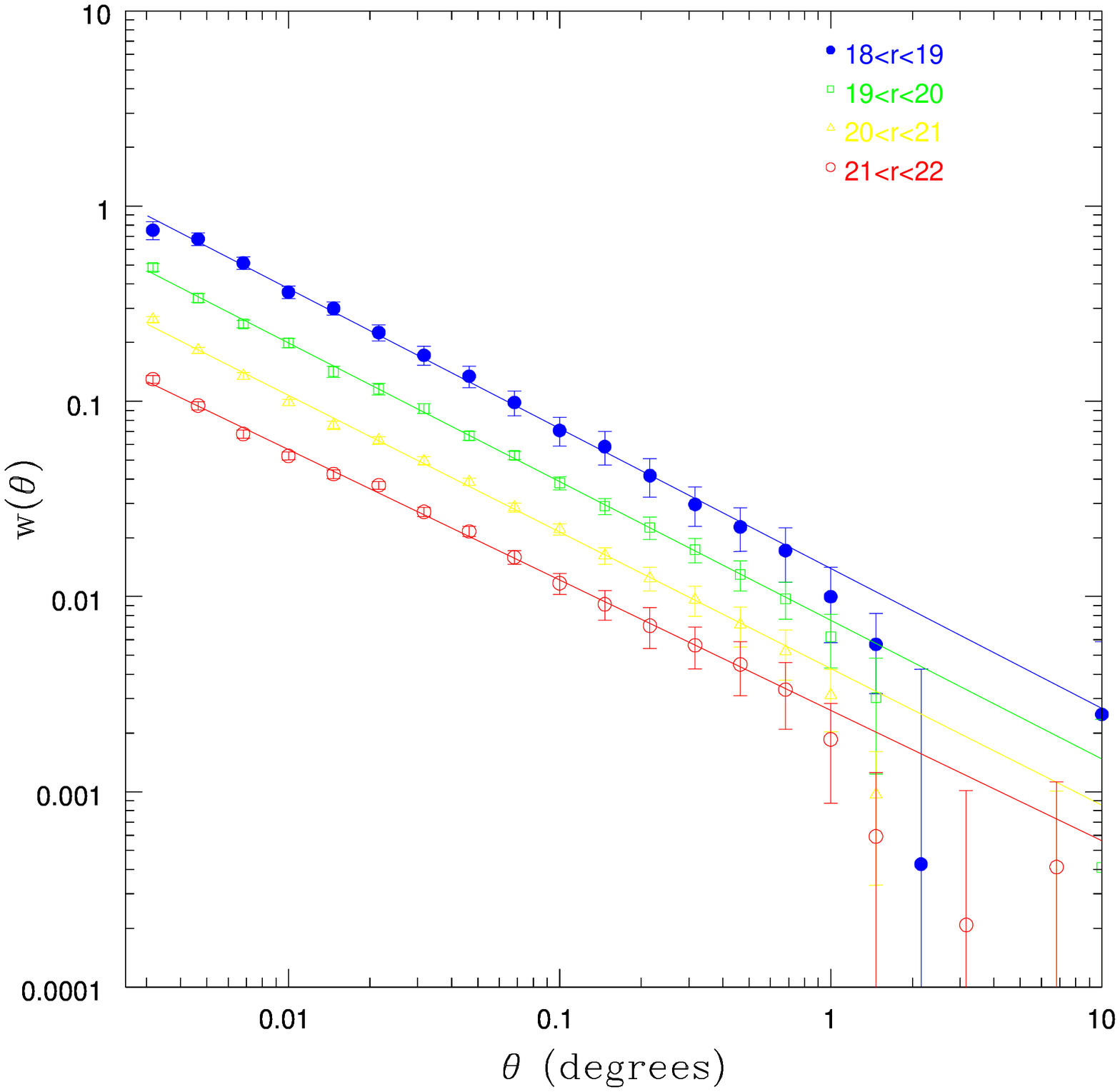}{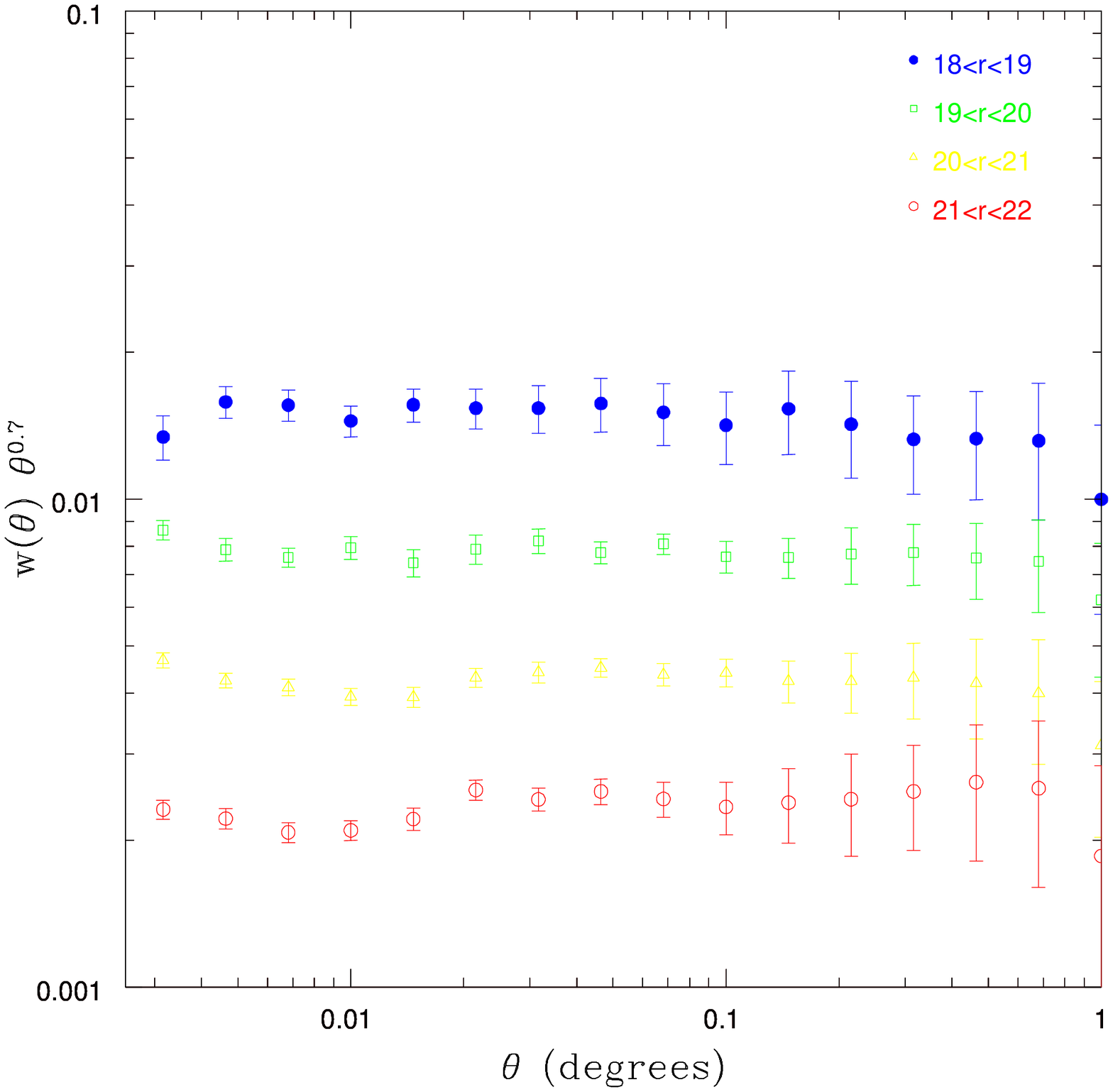}
\label{wm}
\caption{The angular correlation function from the SDSS as a function
of magnitude. Figure 2a shows the correlation function in the
magnitude intervals $18<r^*<19$, $19<r^*<20$, $20<r^*<21$ and
$21<r^*<22$ . The fits to these data, over angular scales of 1' to
30', are shown by the solid lines. All fits use the full covariance
matrix derived from the mock catalogs described in Scranton \etal\
(2001). Figure 2b shows the correlation function for the same
magnitude intervals but with a representative power-law model of
$\theta^{-0.7}$ scaled out.}
\end{figure}

\begin{table}[tb]\footnotesize
\caption{\label{powerlawfits}}
\begin{center}
{\sc Power-law Model Fits\\}
\begin{tabular}{cccc}
\hline\hline
$r^*$ & $\log_{10} A_w$ & $1-\gamma$ & $\chi^2/$dof\\
\hline
18--19 & $-1.90$  $\pm$   0.16  &  $-0.744$  $\pm$   0.038 & 2.41\\
19--20 & $-2.13$  $\pm$   0.13  &  $-0.722$  $\pm$   0.031 & 1.01\\
20--21 & $-2.35$  $\pm$   0.11  &  $-0.700$  $\pm$   0.026 & 1.49\\
21--22 & $-2.58$  $\pm$   0.11  &  $-0.698$  $\pm$   0.026 & 1.23\\
\hline
\end{tabular}
\end{center}
NOTES.---%
Power-law models were fit over a range of 1' to 30', using  
covariance matrices derived from mock catalogs described in
\protect\citet{scranton01}.
\end{table}

Figure 2b shows the angular correlation functions with a
representative $\theta^{-0.7}$ power-law model scaled out. The data
for the fainter magnitude bins clearly deviate from a power-law on
scales smaller than 1 arcminute ($\simeq$ 0.017 degrees). Power-law
fits extending to scales below 1' typically have a reduced $\chi^2$ on
the order of 4 for $r^*>20$, which indicates that such models are not
acceptable fits to the data. We regard this as a preliminary result,
given the difficulties in modeling the covariance matrix to the
required accuracy, due to the fact that the error distribution on
small scales might not be described accurately by a Gaussian, and look
forward to refining this measurement as more of the SDSS survey volume
is completed.

In Figure 3 we show a comparison of the amplitude of the correlation
function at 1$^\circ$, $A_{\omega}$, between the results from the SDSS
data with those derived from earlier deep R band surveys.  The data
shown are derived from the surveys of Stevenson {\it et al.}  (S85;
1985), Couch {\it et al.}  (CJB; 1993), Roche {\it et al.} (R93;
1993), and Hudon \& Lilly (HL96; 1996) and have been converted to the
SDSS $r^*$ magnitude system.  Within the uncertainties in the data the
decrease in the observed correlation function amplitude as a function
of magnitude is consistent with the earlier surveys. It should be
noted, however, that the errors presented here for the SDSS data are
derived from the full covariance matrix whereas the errors for the
majority of the other surveys assume that the points are uncorrelated,
which tends to cause an underestimate of the errors especially as the
fractional width of the angular bins is decreased.

\begin{figure}[h]
\centerline{\psfig{file=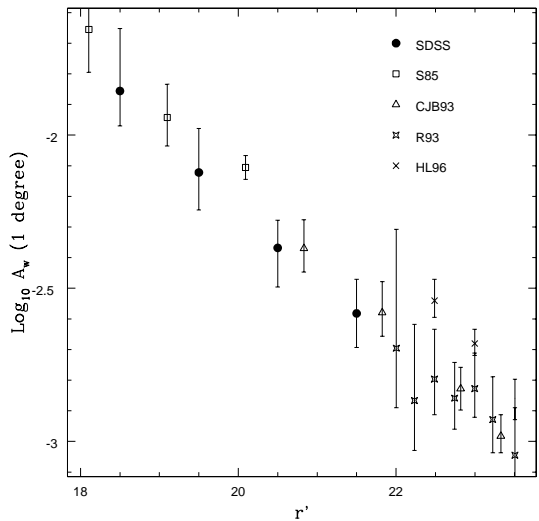,width=0.45\textwidth}}
\label{wm}
\caption{A comparison of the 1$^\circ$ amplitude of the angular
2-point correlation function between large CCD and photographic-plate
surveys. The solid circles show the amplitude of the correlation
function, measured from the SDSS data, as a function of magnitude
interval. The surveys corresponding to the open symbols are described
in the text. Each of the data points corresponds to a one magnitude
interval centered on the point. It should be noted that the error bars
on the SDSS data points are derived from power-law fits using the full
covariance matrix. As noted in the text failure to apply the full
covariance matrix when fitting the highly correlated data will lead to
a severe underestimate of the errors on the fitted parameters.}
\end{figure}

\section{Discussion and Conclusion}

The SDSS represents the first, dedicated, wide-field CCD survey
undertaken for the purpose of characterizing the statistical
properties and clustering of galaxies within the low redshift
Universe. The photometric calibration of the imaging data is currently
accurate to approximately 3\% across the length of the 752 and 756
scans. On completion of the SDSS it is expected that the zero point
calibration across the full survey will be accurate to $\approx
2$\%. Controlling these systematic uncertainties will be crucial to
the success of the large-scale structure analyses of the SDSS.

In this short communication we report some of the first measurements
of the angular clustering for these data.  These initial analyses of
the commissioning data taken in March 1999 show that the shape and
amplitude of the correlation function are consistent with published
wide-field photographic-plate based surveys (scaled to the depth of
the SDSS data). This has also been shown to be true for an independent
subset of the SDSS EDR when galaxies are selected to match the
observed surface density of the APM survey (Gazta\~naga 2001). We find
that the correlation function can be well described by a power law for
scales $1'< \theta < 30'$. The amplitude of the correlation function
decreases as a function of magnitude which is also in good agreement
with deep photographic-plate galaxy surveys and smaller CCD based
photometric surveys. The consistency between this small subset of the
total SDSS data set (we have analyzed less than 2\% of the final areal
coverage) and existing surveys is encouraging and demonstrates the
impact that the full data set will have for measuring galaxy
clustering at low redshift.

On small scales, $\theta < 1$ arcminute, we find that the correlation
function is not accurately described by the same power-law that is
fitted to the larger angular data points. Whether this deviation has a
physical interpretation, arises from low-level systematics present
within the data or is an artifact of the covariance matrix that we
apply (i.e.\ the covariance matrix derived from mock catalogs does not
accurately describe the correlated nature of the data on small scales)
cannot be determined from the current data set. As the SDSS proceeds,
the improvement in the calibration of the data and the increasing
amount of sky surveyed should enable these questions to be addressed.

Beyond the scope of this first generation of clustering papers, but of
importance for subsequent analyses of the SDSS data, is the use of the
colors to estimate the redshifts and spectral types of the
galaxies. The addition of this dimension will also enable the galaxy
samples to be selected based on their restframe properties. This
additional control on sample selection will allow the clustering
analyses to be expressed in terms of the physical properties of the
galaxies rather than purely observational parameters.

\acknowledgments 

The Sloan Digital Sky Survey (SDSS) is a joint project of The
University of Chicago, Fermilab, the Institute for Advanced Study, the
Japan Participation Group, The Johns Hopkins University, the
Max-Planck-Institute for Astronomy (MPIA), the Max-Planck-Institute
for Astrophysics (MPA), New Mexico State University, Princeton
University, the United States Naval Observatory, and the University of
Washington. Apache Point Observatory, site of the SDSS telescopes, is
operated by the Astrophysical Research Consortium (ARC).

Funding for the project has been provided by the Alfred P. Sloan
Foundation, the SDSS member institutions, the National Aeronautics and
Space Administration, the National Science Foundation, the
U.S. Department of Energy, the Japanese Monbukagakusho, and the Max
Planck Society. The SDSS Web site is http://www.sdss.org/.

AJC acknowledges partial support from NSF grant AST0096060 and a NSF
CAREER award AST9984924. DJE was supported by NASA through Hubble
Fellowship grant \#HF-01118.01-99A from the Space Telescope Science
Institute, which is operated by the Association of Universities for
Research in Astronomy, Inc, under NASA contract NAS5-26555.

\end{document}